\documentclass[conference]{IEEEtran}

\def\ps@headings{%
\def\@oddhead{\mbox{}\scriptsize\rightmark \hfil \thepage}%
\def\@evenhead{\scriptsize\thepage \hfil \leftmark\mbox{}}%
\def\@oddfoot{}%
\def\@evenfoot{}}
\makeatother
\usepackage{booktabs} 
\usepackage{url}
\usepackage{svg}
\usepackage{etoolbox}
\usepackage{cuted}
\usepackage{lipsum}
\usepackage{amssymb}
\usepackage{array}
\usepackage{amsmath}
\usepackage{tikz}
\usepackage{color}
\usepackage{dirtytalk}
\usepackage{algpseudocode}
\usepackage{graphicx}
\usepackage{mdframed}
\usepackage{tcolorbox}
\usepackage{comment}
\usepackage{cite}
\usepackage{amsmath,amssymb,amsfonts}
\usepackage{textcomp}
\usepackage{url}
\usepackage{etoolbox}
\usepackage{cuted}
\usepackage{lipsum}
\usepackage{array}
\usepackage{color}
\usepackage{dirtytalk}
\usepackage{algpseudocode}
\usepackage[utf8]{inputenc}
\usepackage{amsmath,amsthm}
\usepackage{multirow}
\usepackage[ruled,vlined]{algorithm2e}
\usepackage{fancyvrb}
\usepackage{pifont}
\usepackage{booktabs,makecell}

\usepackage{siunitx}

\theoremstyle{definition}

\theoremstyle{plain}

\ifCLASSINFOpdf
\else
\fi
\begin{document}

\title{P-CFT: A Privacy-preserving and \\Crash Fault Tolerant Consensus Algorithm \\for Permissioned Blockchains}




\author{\IEEEauthorblockN{Wanxin Li\IEEEauthorrefmark{1}~~~~Collin Meese\IEEEauthorrefmark{1}~~~~Mark Nejad\IEEEauthorrefmark{1}~~~~Hao Guo\IEEEauthorrefmark{2}}
\IEEEauthorblockA{\IEEEauthorrefmark{1}
Department of Civil and Environmental Engineering, University of Delaware, U.S.A.\\\IEEEauthorrefmark{2}School of Software, Northwestern Polytechnical University, Taicang Campus, China.\\
\{wanxinli,cmeese,nejad\}@udel.edu, haoguo@nwpu.edu.cn}\\
}


\maketitle

\begin{abstract}


Consensus algorithms play a critical role in blockchains and directly impact their performance. During consensus processing, nodes need to validate and order the pending transactions into a new block, which requires verifying the application-specific data encapsulated within a transaction. This exposes the underlying data to the consensus nodes, presenting privacy concerns. Existing consensus algorithms focus on realizing application security and performance goals, but lack privacy-by-design properties or are resource-heavy and intended for securing permissionless blockchain networks. In this paper, we propose P-CFT, a zero-knowledge and crash fault tolerant consensus algorithm for permissioned blockchains. The proposed consensus algorithm provides inherent data privacy directly to the consensus layer, while still providing guarantees of crash fault tolerance. We conduct experiments using the Hyperledger Ursa cryptographic library, and the results show promise for integrating P-CFT into existing permissioned blockchain systems requiring privacy-preserving and crash fault tolerant features.


\end{abstract}



%

\section{Introduction}
\label{sec:introduction}

The consensus algorithm is a key component of blockchain systems. Geographically dispersed nodes utilize it to reach agreements on the order of transactions to be included in the next block. Additionally, the consensus process ensures the security of the blockchain and the integrity of the associated data by validating all new transactions before they are committed to the global ledger and replicated on all participating nodes. For example, in the Bitcoin \cite{nakamoto2008bitcoin} system, each new data block contains a hash reference to the preceding block, and the consensus process protects the integrity of the previous data by guaranteeing new blocks contain the correct reference hash. 

The choice of consensus algorithm is largely influenced by the application requirements and the type of blockchain, either permissionless or permissioned. 
In permissioned blockchain systems, membership and data access permissions are maintained by a consortium of one or more organizations, and consensus algorithms have the flexibility to relax some security assumptions of permissionless blockchains in favor of better performance \cite{9137695}. A primary distinction between different permissioned consensus algorithms is their level of fault tolerance, which greatly impacts the system's resilience. For example, crash fault tolerance (CFT) provides the guarantee for reaching consensus in scenarios where some components have failed or communication errors occur, and popular consensus algorithms such as Raft \cite{8666147} and Paxos \cite{lamport2001paxos} can provide this feature. 


However, while many of the proposed and widely used consensus algorithms focus on meeting the application security and performance goals, there are few consensus schemes that address the issue of data privacy in the consensus layer of blockchains. During consensus processing, nodes need to validate and order the pending transactions into a new block, which requires verifying the application-specific data encapsulated within the transaction. This exposes the underlying data to the consensus nodes, presenting privacy concerns for certain application areas, such as healthcare, where regional privacy regulations must be enforced and otherwise may hinder the practicality of deploying beneficial blockchain-based systems. 

\subsection{Contributions and Organization}
In this paper, we address the issue of data privacy within the blockchain consensus layer by proposing a novel privacy-preserving and crash fault tolerant consensus algorithm designed for permissioned blockchain networks. The proposed consensus algorithm integrates zero-knowledge proof (ZKP) for transaction data directly into the consensus processing, ensuring data privacy while still providing fast performance. Previous works have focused on providing privacy to blockchain data either within the context of permissionless blockchain networks or outside of the consensus process \cite{ben2014succinct,bunz2018bulletproofs,ben2018scalable}. In contrast to previous work, we propose P-CFT: a privacy-by-design and crash fault tolerant consensus protocol for permissioned blockchains which can provide ZKP-based privacy to transaction data directly within the consensus layer. 

The proposed consensus algorithm leverages the trust assumptions within permissioned blockchain systems to provide faster prover time while maintaining an acceptable verifier time. Additionally, we conduct a theoretical evaluation of the proposed consensus algorithm including its correctness proof, crash fault tolerance and communication complexity. Using the Hyperledger Ursa cryptographic library, we implement the consensus protocol and conduct extensive experimentation to quantify and compare its performance with other state-of-the-art zero knowledge protocols for blockchain.

The rest of the paper is organized as follows: In section II, we discuss the related work, and also provide preliminary knowledge about permissioned blockchain and zero-knowledge proof to lay the foundation for future sections. Section III presents the detailed consensus construction. Next, we give the correctness proof, and analyze the crash fault tolerance and the communication complexity in Section IV. Then, in Section V, we implement the proposed consensus algorithm and compare its performance with other related approaches. Lastly, we provide concluding remarks and possible future work in Section VI.

\section{Related Work And Preliminaries}
\subsection{Related Work}
Blockchain is a promising decentralized network technology that eliminates the need for a central server and can offer a high level of data immutability, integrity, security and provenance. Consequently, blockchain technology has received much research attention in many areas such as finance \cite{9020009, 9003367}, healthcare \cite{9169395,9507353} and transportation \cite{li2020blockchain, guo2020proof, li2021icbc} as a platform to support new decentralized applications (dApps) in the absence of a trusted third party. Blockchain technology and its associated applications can extract much benefit from a consensus algorithm which inherently preserves data privacy. 


Recently, there have been some attempts to propose zero-knowledge-proof (ZKP) protocols designed to work within existing cryptocurrency systems, in order to provide privacy for the underlying data within blockchain transactions and smart contracts. Zero-knowledge succinct non-interactive arguments of knowledge (zk-SNARK) \cite{ben2014succinct} was the first proposed ZKP-based protocol to be successfully integrated into a production-grade blockchain system, Zcash \cite{hopwood2016zcash}, and it provides privacy for financial transactions on the PoW-based ledger. Bulletproofs were later released and implemented in Monero cryptocurrency, enables proving that a committed value is in a range using a logarithmic number of field and group elements \cite{bunz2018bulletproofs}. In the Ethereum blockchain \cite{etherwhitepaper}, zero-knowledge scalable transparent arguments of knowledge (zk-STARK) \cite{ben2018scalable} is being explored as a way to provide transaction data privacy to its existing permissionless blockchain system. A key distinction between zk-SNARK and zk-STARK protocols is that zk-STARK removes the need for a trusted setup process, which eliminates the undesirable trust assumptions inherent to the zk-SNARK protocol. 

However, the existing zero-knowledge protocols have been designed primarily for cryptocurrency trading and decentralized finance applications in permissionless blockchains, and their rigorous security and cryptography assumptions require significantly more complex protocol designs, which increases computational complexity and the resulting proof latency and size. On the other hand, permissioned blockchain networks benefit from relaxed security assumptions due to their properties of administrator-controlled membership and programmable access controls, allowing for different trust assumptions between participants. 
Given these differences, permissioned blockchain applications could benefit from a consensus protocol designed specifically with privacy in mind, which can provide the property of privacy-by-design for transaction data. This motivated us to research and propose the P-CFT, which inherently preserves data privacy while also offering crash fault tolerance property in consensus layer for permissioned blockchain networks. 

\subsection{Permissioned Blockchain}
Blockchain comes in two primary varieties: permissionless and permissioned \cite{miller2019permissioned}. In the permissionless case (e.g., Bitcoin), membership is entirely open and anyone can join the network and view all of the transactions. In contrast, a permissioned blockchain is a closed membership network, where a consortium of one or more entities will make collaborative decisions about membership, data access controls and governance policies. In permissioned blockchain, anyone who is interested in validating transactions or viewing data on the network needs to get approval from a central authority. This is useful for companies, banks, and institutions that are comfortable to comply with the regulations and are very concerned about having complete control of their data. Due to the ability to control membership, permissioned blockchain systems can utilize lighter-weight consensus algorithms than their permissionless counterparts. Furthermore, programmable access controls can be defined within permissioned blockchain systems, providing fine-grained control for on-chain data. These properties make permissioned blockchain technology more attractive for certain applications, where high transaction throughputs with low latencies are required. 


\subsection{Zero-knowledge Proof}
Zero-knowledge proof (ZKP) was proposed in 1989 by Goldwasser, Micali, and Rackoff~\cite{goldwasser1989knowledge}.
In cryptography, a ZKP protocol is a method by which a prover can convince a verifier that he/she knows a secret message $m$, without conveying any information, apart from the fact that the prover knows the secret message $m$~\cite{wiki:zkp}. 
A ZKP of knowledge is a special case when the statement consists only of the fact that the prover possesses the secret information~\cite{wiki:zkp}. Based on the frequency of communications between the prover and the verifier, there are two ZKP schemes: Interactive ZKP and Non-interactive ZKP schemes. 
A zero-knowledge proof must satisfy three properties:

\begin{itemize}
    \item Completeness: If the statement is true, the honest verifier will be convinced of this fact by an honest prover.
    \item Soundness: There is no such prover that can convince an honest verifier if he/she does not compute the results correctly.
    \item Zero-knowledge: The proof of knowledge can be simulated without revealing any secret information, which means that no verifier learns anything other than the fact that the statement is true.
\end{itemize}

The inherent properties of ZKP can be leveraged to ensure data input validity for blockchain transactions without revealing any of the sensitive information during the consensus process.

\section{Consensus Design}
\label{sec: ZK-BFT}

\begin{figure}[bp]
\centering
\includegraphics[width=0.45\textwidth]{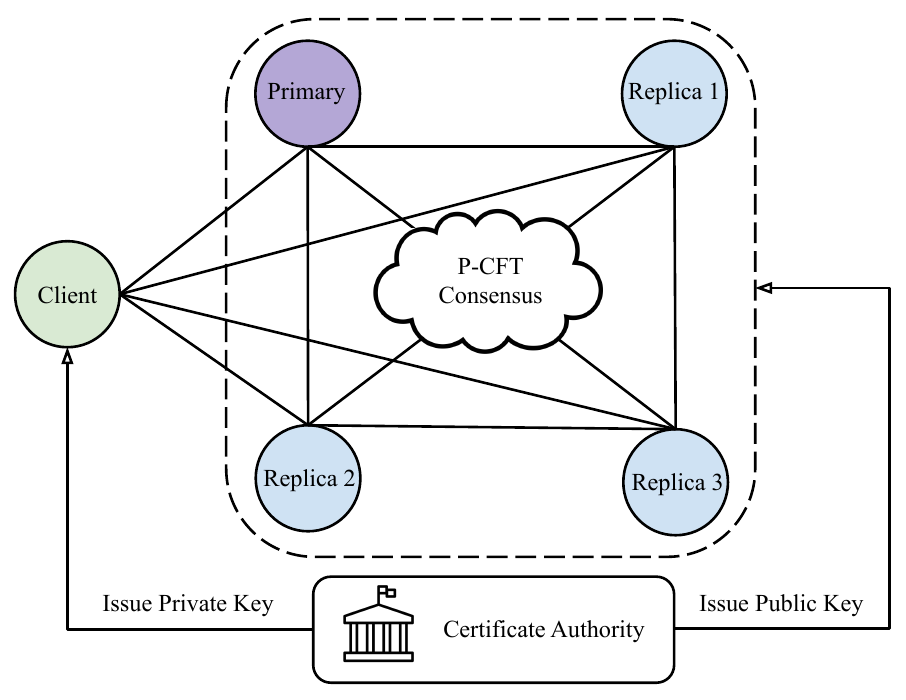}
\caption{Overview of the proposed P-CFT consensus. (Note that we give one client node and four consensus nodes as an example here, the consensus proposed in this paper can be extended to arbitrary number of nodes.)}
\label{fig:overview}
\end{figure}

In this section, we present the detailed construction of the proposed P-CFT consensus. By referring to Figure \ref{fig:overview}, the P-CFT consensus includes the certificate authority and three types of nodes, which are defined as follows:


\begin{itemize}
    \item Certificate Authority (CA): A certificate authority certifies the ownership of clients' digital assets by issuing key pairs. Private keys are sent to each client for generating zero-knowledge proofs, and public keys are sent to the primary node and replica nodes for verification purposes.

    \item Client Node: Client nodes are responsible for generating zero-knowledge proofs and sending transaction requests.
    
    \item Primary Node: Primary node is a healthy leader that is responsible for voting on transactions, building and publishing blocks. Each consensus-reaching process has one and only one primary node.
    
    \item Replica Node: Replica nodes are responsible for voting on transactions and the leader's health.
    
\end{itemize}

In our design, consensus nodes include both primary node and replica nodes. As shown in Figure \ref{fig:ZK-BFT}, the proposed consensus algorithm is constituted by the following steps: the certificate authority issues key pairs for clients and consensus nodes; the client sends the request; the primary node forwards the request; all consensus nodes execute \textit{Verify} process; the client receives the response that the consensus is reached. Together, they form the core of the proposed P-CFT consensus algorithm. Given the total number of $N$ consensus nodes, the consensus is technically reached when the client receives at least $(N-1)/2+1$ replies from the consensus node group. The proposed consensus process is explained in detail as follows:


\begin{figure}[t]
\centering
\includegraphics[width=0.48\textwidth]{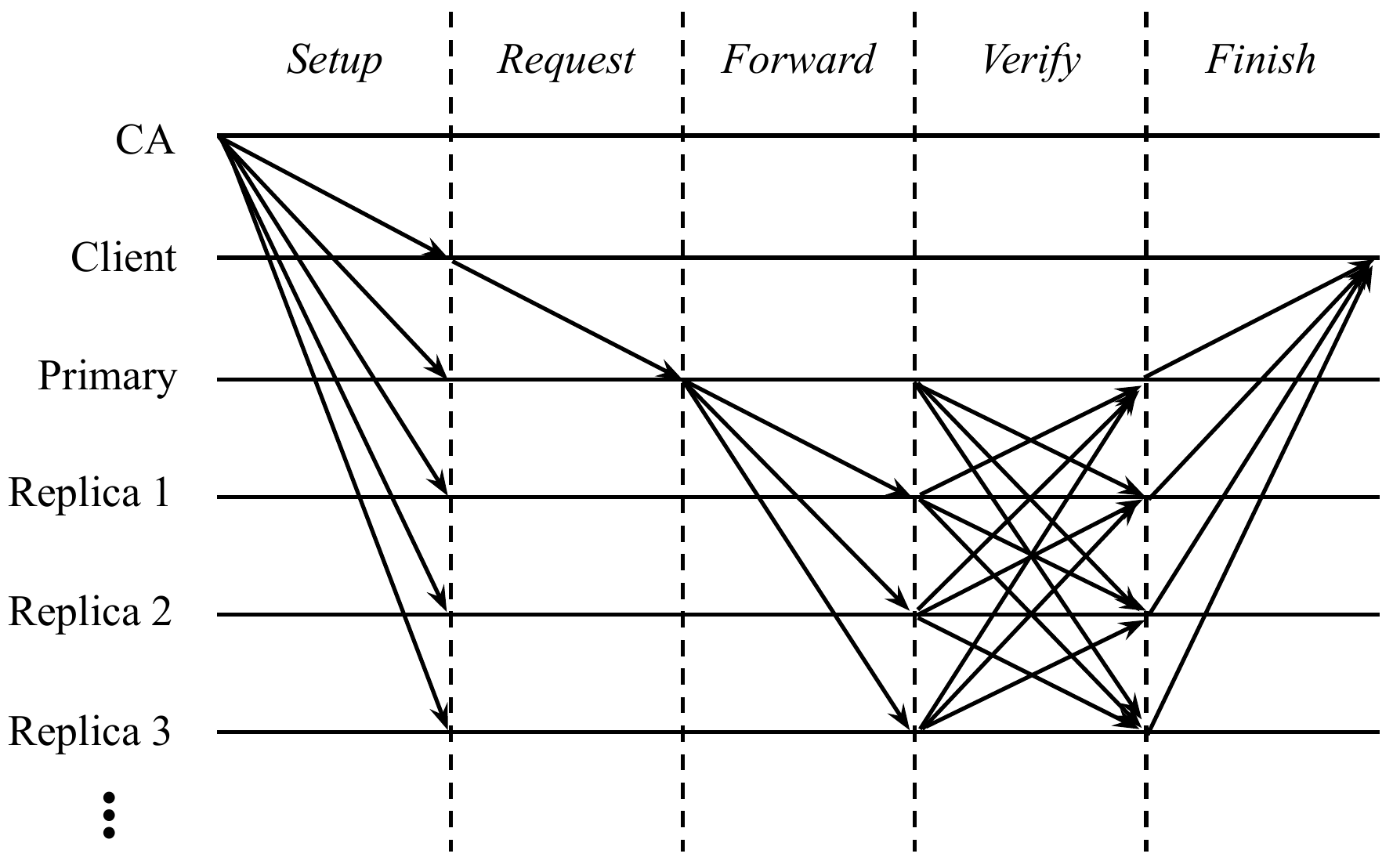}
\caption{The proposed P-CFT consensus processing.}
\label{fig:ZK-BFT}
\end{figure}

\vspace{0.1cm}
\subsubsection{Setup} Inherent to a zero-knowledge proof scheme, in order to prevent the prover from cheating and generating fake proofs, the verifier must know exactly what is being proven. In this process, the certificate authority runs Algorithm \ref{alg: zkp-keygen} to issue key pairs for transaction message (e.g., digital assets). The \textit{Setup} process only needs to be carried out once, and the certificate authority sends the setup message $<SETUP, id, sk>$ to the client node and sends setup messages $<SETUP', id, pk>$ to the consensus nodes. The $id$ is an identifier to indicate the original message $m$, $sk$ represents the private key and $pk$ represents the public key.

\begin{algorithm}[ht]
\label{alg: zkp-keygen}
\SetAlgoLined
\LinesNumbered
\SetKwInOut{Input}{Input}
\SetKwInOut{Output}{Output}
\Input{message $m$}
\Output{private key $sk$, public key $pk$}
The certificate authority selects a random $a \in \mathbb{Z}_p$\ for message $m$\;
The certificate authority saves the private key as $sk = a$\;
The certificate authority computes the public key as $pk = g^{sk} \in \mathbb{G}$\;
The certificate authority returns $sk$ and $pk$\;
\caption{\textit{KeyGen}}
\end{algorithm}

\vspace{0.1cm}
\subsubsection{Request} In this process, the client runs Algorithm \ref{alg: zkp-proofgen} to generate the one-time zero-knowledge proof $\delta$ based on the original message $m$, and starts to send a request $<REQUEST, id, h, \delta>$ to the system. The $h$ represents the hash digest of the original message $m$. 

\begin{algorithm}[ht]
\label{alg: zkp-proofgen}
\SetAlgoLined
\LinesNumbered
\SetKwInOut{Input}{Input}
\SetKwInOut{Output}{Output}
\Input{message $m$, private key $sk$}
\Output{one-time zero-knowledge proof $\delta$}
The client computes a hash digest $h$ based on transaction message $m$
, as $h = H(m)$\;
The client generates the one-time zero-knowledge proof $\delta = {h}^{sk} \in \mathbb{G}$\;
The client returns $\delta$\;
\caption{\textit{ProofGen}}
\end{algorithm}



\vspace{0.1cm}
\subsubsection{Forward} In the third process, The primary node publishes a new block and broadcasts the client's request in messages $<FORWARD, id, h, \delta, v>$ to the other replica nodes. The $v$ represents the view number.

\vspace{0.1cm}
\subsubsection{Verify} Replica nodes receive the forwarded messages and verify the following: (1) The node is currently in the view $v$; (2) The node does not have other \textit{Forward} messages on the same page (view $v$, message identifier $id$). In other words, there is not another set of ($h', \delta'$) that shares the same message identifier $id$ with the set of ($h, \delta$) in the current view $v$; (3) The node runs Algorithm \ref{alg: zkp-proofverify} to verify the authenticity of the one-time zero-knowledge proof $\delta$ without accessing the original message $m$. After the verification is successfully done, replica nodes send out the corresponding verification messages $<VERIFY, id, h, \delta, v, i, r>$ to the other consensus nodes. The $i$ represents the identity of the replica node and $r$ is a Boolean value ($true$ or $false$) that indicates its verification result.

\begin{algorithm}[ht]
\label{alg: zkp-proofverify}
\SetAlgoLined
\LinesNumbered
\SetKwInOut{Input}{Input}
\SetKwInOut{Output}{Output}
\Input{one-time zero-knowledge proof $\delta$, public~key~$pk$, generator $g$, hash~digest~$h$}
\Output{verification result $r$}
the consensus node checks
\eIf{$e(\delta, g) == e(h, pk)$}{$r = true$ \;}{$r = false$ \;}
The consensus node returns $r$ \;
\caption{\textit{ProofVerify}}
\end{algorithm}



\vspace{0.1cm}
\subsubsection{Finish} Each consensus node needs to receive at least $(N-1)/2$ \textit{Verify} messages from other consensus nodes (a total of $(N-1)/2+1$ including its own) and validates if the $id, h, \delta, v$ of these verification messages are all consistent.
Then, each node commits the block for which they have the matching \textit{Forward} and at least $(N-1)/2+1$ \textit{Verify} messages. After the block has been successfully committed to the chain, each node will send a \textit{Finish} message to the client that a consensus is reached on its request.



A view is the period of time that a given node is the primary. Therefore, a view change is switching to a different primary node. When a replica node determines that the current view $v$ is faulty, such as the primary node sent an invalid message or did not produce a valid block in time, it will broadcast a view change request for $v + 1$ to the other nodes in the network. After receiving the request, the other nodes will verify it by communicating with the current primary node. If the primary is indeed faulty, all non-faulty nodes will broadcast the confirmation messages for view change. 

\section{Discussion and Analysis}

\subsection{Correctness Proof}
\noindent
\textbf{Proposition 1.} \textit{The proposed P-CFT consensus can correctly verify the transaction request without revealing the original message $m$.}
\vspace{1mm}

Assume that a client generates the zero-knowledge proof $\delta$ for the message $m$ and sends this proof $\delta$ to the primary node through the transaction request. We prove that the proof $\delta$ is validated by the Algorithm \ref{alg: zkp-proofverify}. First, the public key $pk$ is computed as:
\begin{equation}
    pk = g^{sk},
\end{equation}

Then, a one-time zero-knowledge proof is generated as:
\begin{equation}
\label{equation-generate}
   (m, sk) \longrightarrow \delta = {H(m)}^{sk} = h^{sk},
\end{equation}

Next, verifying the proof $\delta$ is done by checking that iff:
\begin{equation}
\label{equation-verify}
   e(\delta, g) = e(h, pk),
\end{equation}

Now, we prove the Equation \ref{equation-verify} based on bilinear pairing property: 
\begin{equation}
\begin{split}
    e(\delta, g) &= e(h^{sk}, g) \\
    &= e(h, g^{sk}) \\
    &= e(h, pk).
\end{split}    
\end{equation}

\noindent
\textbf{Bilinear Pairing Property:} \textit{Let $\mathbb{G}$ be a multiplicative cyclic group of prime order $p$ with generator $g$. Let $e : \mathbb{G} \times \mathbb{G} \rightarrow \mathbb{G}_T$ be a computable, bilinear and non-degenerate pairing into the group $\mathbb{G}_T$. Then, we have $e(x^a, y^b) = e(x, y)^{ab}$ for all $x,y \in \mathbb{G}$ and $a,b \in \mathbb{Z}_p$ because $\mathbb{G}$ is cyclic.}

\subsection{Crash Fault Tolerance}
\noindent
\textbf{Proposition 2.} \textit{The proposed P-CFT consensus can tolerate up to $f$ nodes that fail for communication ($f = (N-1)/2$). In other words, P-CFT consensus can offer $1/2$ crash fault tolerance.}
\vspace{1mm}

Crash fault tolerance is the important resiliency that the distributed system can still correctly reach consensus if certain nodes fail for communication. Given a total number of $N$ consensus nodes ($N = 2f + 1$), each node needs to receive at least $f+1$ \textit{Verify} messages from other nodes to successfully process the \textit{Finish} phase resulting in $1/2$ crash fault tolerance. 

\subsection{Communication Complexity} 
\noindent
\textbf{Proposition 3.} \textit{Given the total node number $N$ including 1 primary node and $n$ replica nodes ($N=n+1$), the communication complexity to reach consensus in the proposed P-CFT consensus is $O(N^2)$.}

\vspace{1mm}
The communication complexity of the proposed consensus algorithm is in the order of $O(N^2)$ because of the peer-to-peer and all-to-all communications from the \textit{Verify} phase, as shown in Figure \ref{fig:ZK-BFT}. More specifically, in the \textit{Verify} phase, after the verification is done, each replica node sends out the result to the other consensus nodes including the primary node, generating a total of $N*N$ messages. 

\begin{table*}[t]
\centering
\caption{Comparisons of the proposed and state-of-the-art ZKP protocols for Blockchain.}
\label{tab: zkp comparison}
\begin{tabular}{|c|c|c|c|c|c|}
\hline
\thead{\textbf{ZKP Protocol}} & \textbf{Blockchain Type} & \textbf{Blockchain Layer} & \textbf{Non-interactive} & \textbf{Prover Time} & \textbf{Verifier Time}\\ \hline
zk-SNARKs \cite{ben2014succinct}  & Permissionless & Data  & \checkmark            & $2,300ms$     & $10ms$ \\ \hline
Bulletproofs \cite{bunz2018bulletproofs} & Permissionless & Data & \checkmark             & $30,000ms$    & $1,100ms$ \\ \hline
zk-STARKs \cite{ben2018scalable} & Permissionless & Application & \checkmark             & $1,600ms$     & $16ms$ \\ \hline
Proposed P-CFT    & Permissioned  & Consensus & \checkmark             & $31ms$        & $214ms$ \\ \hline
\end{tabular}
\end{table*}

\section{Implementation and Comparison}
\label{sec: fs-zkp}
\subsection{Implementation}

\begin{figure}[ht]
\centering
\includegraphics[width=0.4\textwidth]{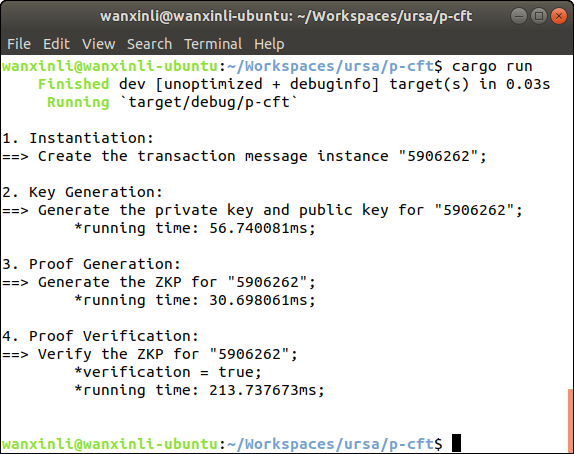}
\caption{Prototype of the zero-knowledge protocol on Hyperledger Ursa.}
\label{fig:fs-zkp}
\end{figure}

We developed the zero-knowledge protocol in the Rust programming language \cite{matsakis2014rust} utilizing the Hyperledger Ursa library \cite{hyperledgerursa} on Ubuntu 18.04 operating system with 2.8 GHz Intel i5-8400 processor and 8GB DDR4 memory. We chose the SHA256 algorithm \cite{rachmawati2018comparative} to generate the one-way hash digest $h$ for message $m$ and the elliptic curve \cite{frey1999tate} for bilinear pairing $e$. 
As shown in Figure \ref{fig:fs-zkp}, the prototype provides the functionalities of instantiation, key generation, ZKP generation and ZKP verification defined in Algorithms \ref{alg: zkp-keygen}, \ref{alg: zkp-proofgen} and \ref{alg: zkp-proofverify}. The average running times of each phase are 57ms, 31ms and 214ms, respectively.

\subsection{Comparison with Other ZKP Protocols}
In this subsection, we compare the performance among the proposed protocol and three state-of-the-art zero-knowledge proof protocols which provide privacy for blockchain technology, including zk-SNARKs \cite{ben2014succinct}, BulletProofs \cite{bunz2018bulletproofs} and zk-STARKs \cite{ben2018scalable}.

The zk-SNARKs was introduced by Bitansky et al. in 2012 \cite{bitansky2012extractable}. The first widespread application of zk-SNARKs was in the Zerocash blockchain protocol, where zero-knowledge cryptography provides the computational backbone by facilitating mathematical proofs that one party has possession of certain information without revealing what that information is \cite{sasson2014zerocash}. Bulletproofs were released in 2018 and later implemented in Monero cryptocurrency, which enables proving that a committed value is in a range using a logarithmic number of field and group elements \cite{bunz2018bulletproofs}. In 2018, the zk-STARKs protocol was introduced \cite{ben2018scalable}, offering faster prover time than zk-SNARKs while also removing the trusted setup process.

\begin{figure}[t]
\centering
\includegraphics[width=0.45\textwidth]{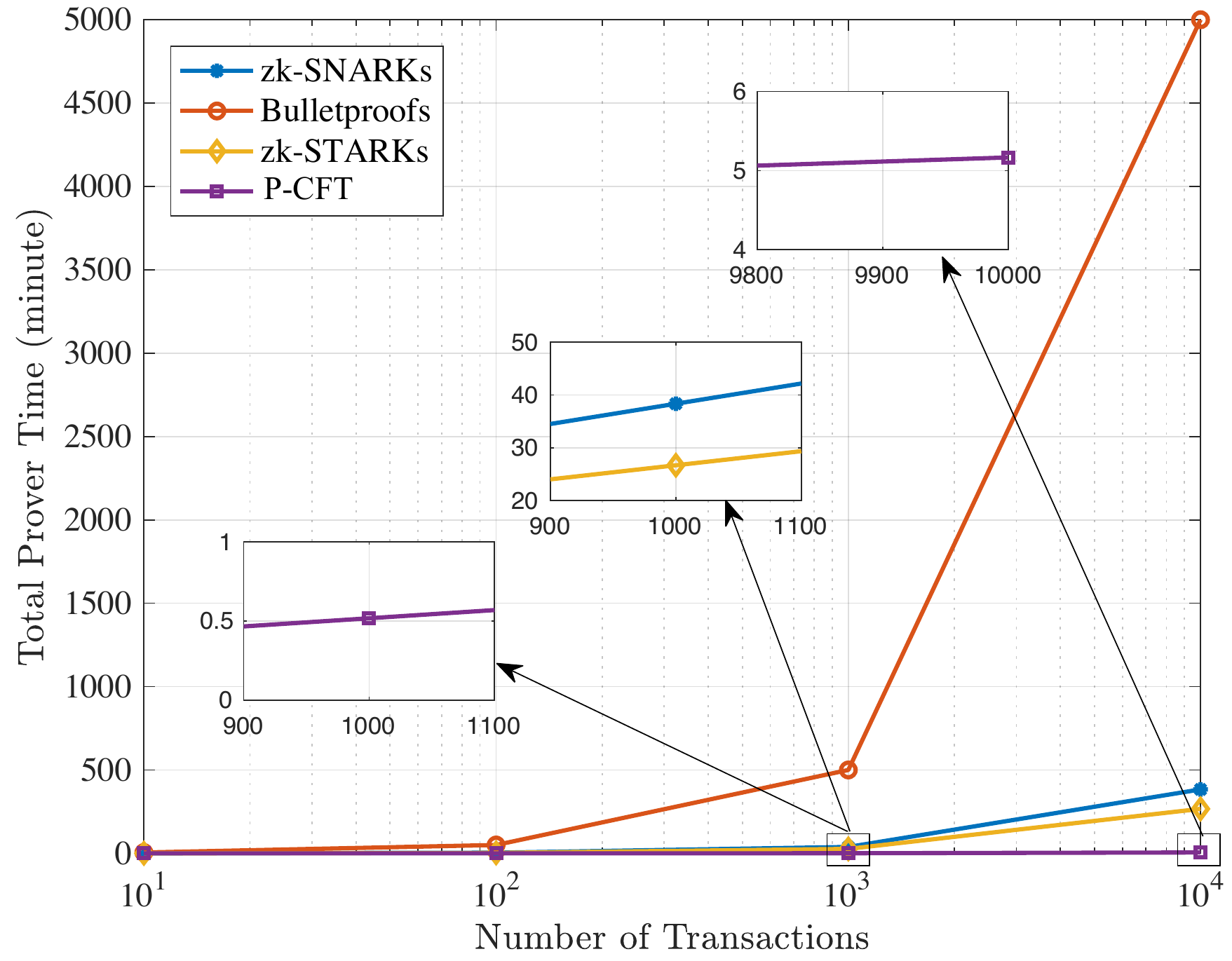}
\caption{Total prover time vs. the number of transactions in comparison among the state-of-the-art and the proposed zero-knowledge protocols.}
\label{fig: provertime}
\end{figure}

\begin{figure}[t]
\centering
\includegraphics[width=0.44\textwidth]{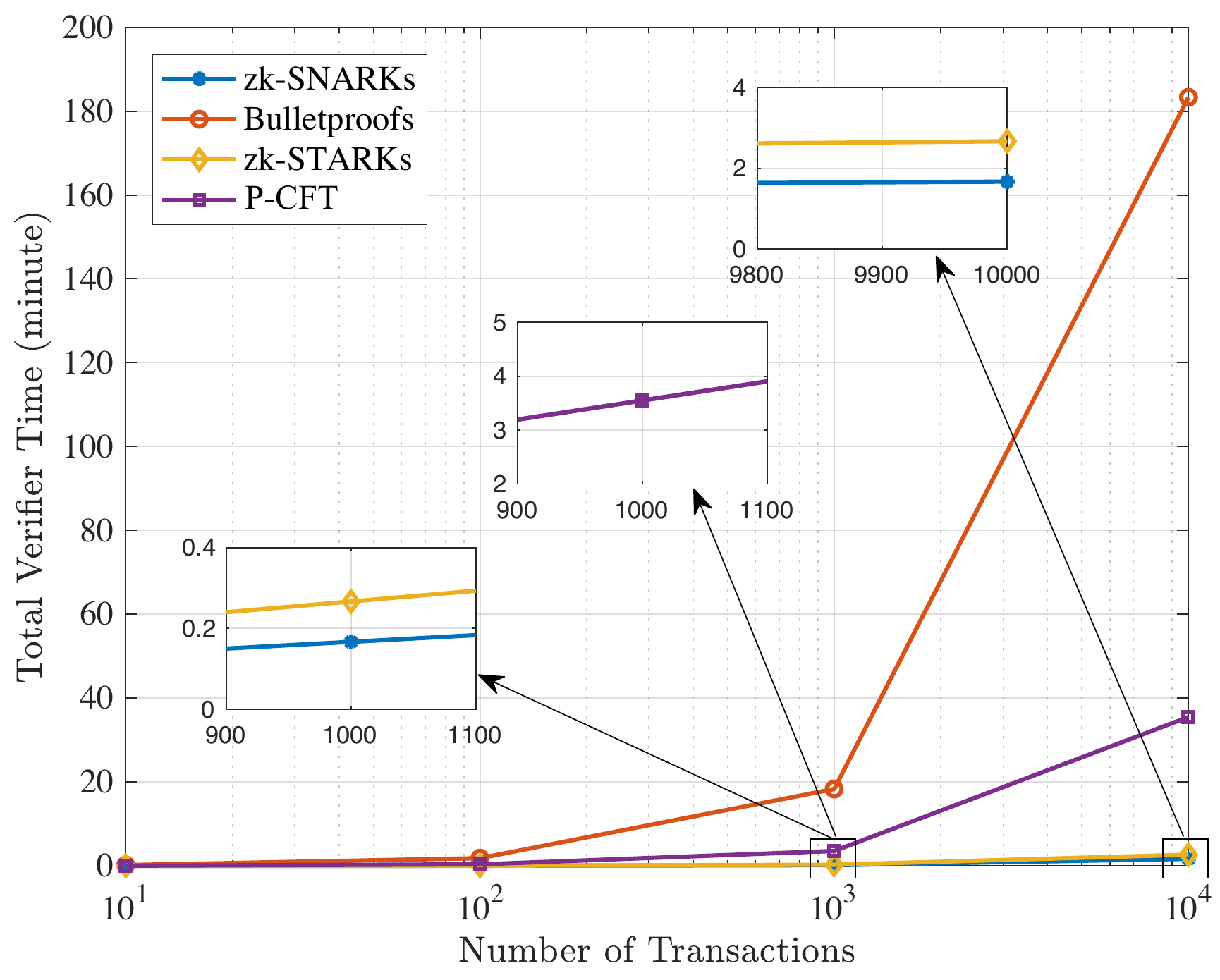}
\caption{Total verifier time vs. the number of transactions in comparison among the state-of-the-art and the proposed zero-knowledge protocols.}
\label{fig: verifiertime}
\end{figure}

As shown in Table \ref{tab: zkp comparison}, we compare the performance of the proposed zero-knowledge protocol with the above-mentioned three state-of-the-art ZKP protocols. Notably, all three of the state-of-the-art ZKP protocols have been integrated into either the data or application layer for cryptocurrencies, while P-CFT is designed for the consensus layer of permissioned blockchain networks. Furthermore, non-interactive zero-knowledge proofs refer to zero-knowledge proofs that require no interaction between the prover and verifier. The proposed zero-knowledge protocol is designed as a non-interactive version of zero-knowledge proof to validate the proof only in one round. Moreover, zk-STARKs is faster than zk-SNARKs and Bulletproofs at the prover level (1.6s), while the protocol is slightly slower than zk-SNARKs at the verifier level. The prover time is significantly decreased in the proposed zero-knowledge protocol, generating the proof in only 31ms. P-CFT also provides an acceptable verifier time for proof validation in 214ms. 

The performance advantage of the proposed consensus algorithm becomes more apparent when we increase the number of transactions to a larger scale. Figure \ref{fig: provertime} shows the total prover time by varying the number of transactions. The proposed P-CFT is able to handle 10,000 transactions in five minutes at the prover level, which saves hundreds of minutes compared to zk-SNARKs and zk-STARKs and thousands of minutes compared to Bulletproofs. Figure \ref{fig: verifiertime} shows the total verifier time by varying the number of transactions. The proposed zero-knowledge protocol can verify 10,000 transactions in 35 minutes at the verifier level, which is slower than zk-SNARKs (1.7 minutes) and zk-STARKs (2.7 minutes) but saves hundreds of minutes compared to Bulletproofs.

\section{Conclusion}
This paper proposes a zero-knowledge and crash fault tolerant consensus algorithm for permissioned blockchains, which brings privacy-by-design directly to the consensus layer. In the theoretical analysis of the proposed P-CFT consensus, we provide proofs for correctness, crash fault tolerance and communication complexity. In order to evaluate the proposed system, we developed the zero-knowledge protocol in the Rust programming language  utilizing the Hyperledger Ursa cryptographic library. The results show that the proposed protocol can provide fast proof generation time, while maintaining a low verification time in comparison with the existing ZKP protocols. 
Consequently, the results demonstrate the feasibility of the proposed approach for providing privacy to the consensus level of existing and production-grade permissioned blockchain networks. For future work, we plan to improve the fault tolerance level of the proposed consensus algorithm to provide resilience against Byzantine attacks.


\ifCLASSOPTIONcompsoc
  \section*{Acknowledgments}
\else
  \section*{Acknowledgment}
This work is partially supported by the Fundamental Research Funds for the Central Universities under the Grant G2021KY05101.
  
\fi

\Urlmuskip=0mu plus 1mu\relax
\bibliographystyle{IEEEtran}
\bibliography{sig.bib}

\end{document}